\documentclass{rspublic}
\usepackage{graphicx}

\begin{document}

\title[Emergent physics: Fermi point scenario]{Emergent physics: Fermi point scenario}

\author[G.E. Volovik]{Grigory Volovik}

\affiliation{Low Temperature Laboratory,
Helsinki University of Technology\\
P.O. Box 2200, FIN-02015 HUT, Finland\\
and\\
L.D. Landau Institute for Theoretical Physics,
Russian Academy of Sciences, Kosygina 2, 119334 Moscow, Russia\\}
\label{firstpage}

\maketitle

\begin{abstract}{Fermi point, emergent physics, cosmological constant}
The Fermi-point scenario of emergent gravity has the following consequences:
 gravity emerges together with fermionic and bosonic matter;
 emergent fermionic matter consists of massless Weyl fermions;
 emergent bosonic matter consists of gauge fields;
 Lorentz symmetry persists well above the Planck energy;
 space-time is naturally 4-dimensional;
 Universe is naturally flat; cosmological constant is naturally small or zero;
 underlying physics is based on discrete symmetries;
 `quantum gravity' cannot be obtained by quantization of Einstein equations;
there is no contradiction between quantum mechanics and gravity;
etc.

\end{abstract}

\section{Introduction}

Astronomical observations suggest the existence of a cosmological constant introduced by Einstein (1917). Its value corresponds to the vacuum energy density
(Bronstein 1933, Zeldovich 1967)  of order $\Lambda_{\rm obs}\sim E_{\rm obs}^4$ with the characteristic  energy scale   $E_{\rm obs}\sim 10^{-3}\;\text{eV}$.
Current theories do not provide any good symmetry explanation for the smallness of this value as compared with naive theoretical estimation suggesting the Planck scale for the vacuum energy: $\Lambda_{\rm theor}\sim E_{\rm P}^4$ with $E_{\rm P}\sim 10^{19}\;\text{GeV}$. This is the so called cosmological constant problem.
Another huge disagreement between the naive expectations and observations  concerns
masses of elementary particles. The naive estimation tells us that these masses should be on the order of Planck energy scale:  $M_{\rm theor} \sim E_{\rm P}$, while the  masses of observed particles are many orders of magnitude smaller being below the electroweak energy scale $M_{\rm obs}<E_{\rm ew}\sim 1$  TeV.  This  is called the hierarchy problem.
There should be a general principle, which could resolve both paradoxes. Here we discuss the principle of emergent physics based on the topology in momentum space. 

The momentum space topology  suggests that both in relativistic quantum field theories and in the fermionic condensed matter there are several universality classes of ground states -- quantum vacua  (Horava 2005). One of them contains vacua with trivial topology, whose fermionic excitations are massive (gapped) fermions. The natural mass  of these fermions is on the order of $E_{\rm P}$.
However, the other classes contain gapless vacua. Their fermionic excitations live either near Fermi surface (as in metals), or near Fermi point (as in superfluid $^3$He-A) or near some other topologically stable manifold of zeroes in the energy spectrum. The gaplessness of these fermions is protected by topology, and thus is not sensitive to the details of the microscopic (trans-Planckian) physics. Irrespective of the deformation of the parameters of the microscopic theory, the value of the gap (mass) in the energy spectrum of these fermions remains strictly zero. This solves the main hierarchy problem: for these classes of  fermionic vacua the masses of elementary particles are naturally small. 

In the emergent physics, the vacuum energy which is relevant for gravity is naturally small. This can be checked on the example of the self-sustained vacua, i.e. the vacua which can be in equilibrium in the absence of environment (Klinkhamer  \& Volovik 2007).  The energy density of such vacua is strictly zero if the vacuum is perfect and is isolated from environment.  This solves the main cosmological constant problem: $\Lambda$ is naturally small.

\section{Fermionic and bosonic content in vacua with Fermi points}

For our Universe, which obeys the Lorentz invariance,  only those vacua  are important that are either Lorentz invariant, or acquire the Lorentz invariance as an effective  symmetry emerging at low energy. This excludes the vacua with Fermi surface and leaves the class of vacua with Fermi point of chiral type, in which fermionic excitations behave as left-handed or right-handed Weyl fermions   (Froggatt \& Nielsen 1991, Volovik 2003), and the class of vacua with the nodal point obeying $Z_2$ topology, where fermionic excitations behave as massless Majorana neutrinos   (Horava 2005, Volovik 2007). General properties of quantum vacua obeying  Lorentz invariance are discussed by Klinkhamer  \& Volovik (2007).

 \subsection{Emergent fermionic matter}

The advantage of the vacua with Fermi points is that practically all the main physical laws (except for quantum mechanics) can be considered as effective laws, which naturally emerge at low energy. This is the consequence of the  so-called Atiyah-Bott-Shapiro construction (Horava 2005),  which leads to the following general form of expansion of the inverse fermionic propagator near the Fermi point:
\begin{equation}
G^{-1}(p_\mu)=e_\alpha^\beta\Gamma^\alpha(p_\beta-p_\beta^{(0)})+~{\rm higher~order~terms}~.
\label{Atiyah-Bott-Shapiro}
\end{equation}
Here $\Gamma^\mu=(1,\sigma_x,\sigma_y,\sigma_z)$ are Pauli matrices (or Dirac matrices in the more general case); the expansion parameters are the vector 
$p_\beta^{(0)}$ indicating the position of the Fermi point in momentum space where the Green's function has a singularity, and  the matrix $e_\alpha^\beta$). This expansion is written for the simplest case of the isolated Fermi point with the elementary  topological charges, i.e. either with $N=+1$ or $N=-1$. The equation (\ref{Atiyah-Bott-Shapiro}) can be transformed to the form 
\begin{equation}
G^{-1}(p_\mu)=ip_0 +N\boldsymbol{\sigma}\cdot{\bf p}+~{\rm higher~order~terms}~,
\label{Weyl}
\end{equation}
where the position of the Fermi point is shifted to $p_\beta^{(0)}=0$; the matrix $e_\alpha^\beta$ is transformed to unit matrix; and $p_\mu=(ip_0,{\bf p})$. This form demonstrates that close to the Fermi point with $N=+1$, the low energy fermions behave as right handed relativistic particles, while the Fermi point with $N=-1$ gives rise to the left handed particles. This scenario agrees with the fermionic content of our Universe, where all the elementary particles -- left-handed and right-handed quarks and leptons -- are Weyl fermions. 

In principle, the infrared divergences may violate the simple pole structure of the propagator in Eq.(\ref{Weyl}), and one will have 
\begin{equation}
G(p_\mu)\propto \frac{-ip_0 +N\boldsymbol{\sigma}\cdot{\bf p}}{\left(p^2+p_0^2\right)^{\gamma}} ~,
\label{WeylUnparticle}
\end{equation}
with  $\gamma\neq 1$.  This modification does not change the topology of the propagator: its topological charge is $N$ for arbitrary parameter  $\gamma$  (Volovik 2007). Fermions without pole in the Green's function occur in condensed matter, in particular in Luttinger Fermi liquids (Giamarchi  2004),  and may also occur in relativistic quantum fields, see e.g.  fermionic unparticles with 
$\gamma=5/2 -d_U$, where $d_U$ is the scale dimension of the quantum field (Georgi 2007, Luo  \&  Zhu 2008). In the Fermi-point scenario, the form of the propagator in Eq.(\ref{WeylUnparticle}) is dictated by topology in momentum space.

 \subsection{Emergent gauge fields}
 
   \begin{figure}
\centerline{\includegraphics[width=1.0\linewidth]{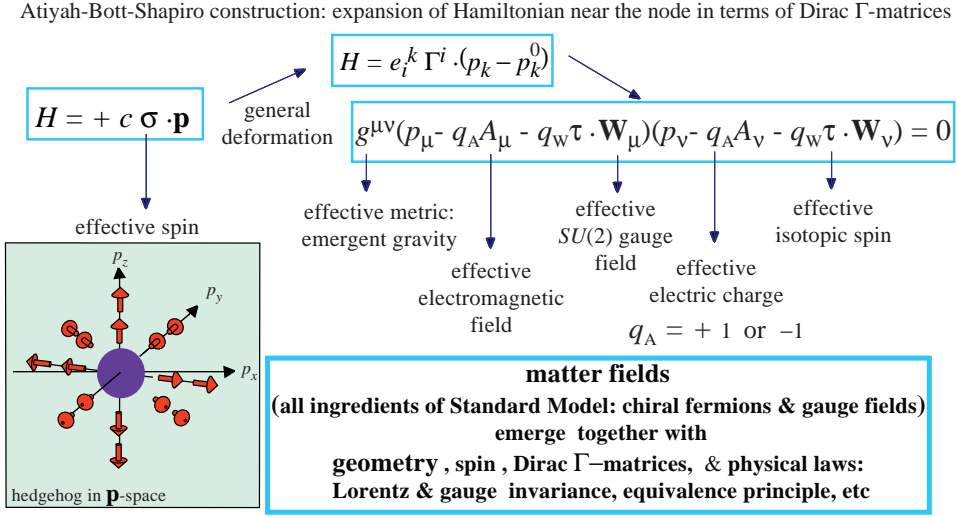}}
\caption{
Fermi point is the topologically stable hedgehog in momentum space. Close to the Fermi point with topological charge $N=+1$, effective spin of the fermionic particle is directed along the momentum, $\boldsymbol{\sigma}\parallel {\bf p}$, forming the hedgehog with spines outward ({\it bottom left}). This means that in the vicinity of such Fermi point, fermions behave as right-handed relativistic particles.  Bosonic quantum fields emerging at low energy correspond to relativistic quantum gauge fields and gravity. Near Fermi point with multiple topological charge $|N|>1$, effective non-Abelian gauge fields emerge. In particular, the $SU(2)$ gauge field ${\bf W}_\mu$ emerges near Fermi point with the topological charge $N=\pm 2$ and with the underlying $Z_2$ discrete symmetry (Sec. \ref{SymmetryTopology}).}  
\label{EmergentPhysics} 
\end{figure}

The Fermi-point scenario gives a particular mechanism for emergent symmetry. The Lorentz symmetry is simply the result of the linear expansion: it becomes better and better when the Fermi point is approached and the non-relativistic higher order terms in Eq.(\ref{Weyl}) may be neglected. This expansion  demonstrates the emergence of the relativistic spin, which is described by the Pauli matrices. It also demonstrates how gauge fields and gravity emerge together with chiral fermions. 
The expansion parameters  
$p_\beta^{(0)}$ and $e_\alpha^\beta$) may depend on the space and time coordinates and they actually represent collective dynamic bosonic fields in the vacuum with Fermi point. 
The vector field  $p_\beta^{(0)}$ in the expansion plays the role of   the effective $U(1)$ gauge field $A_\beta$acting on  fermions. 
For the more complicated Fermi points, the shift $p_\beta^{(0)}$ becomes the matrix field; it gives rise to effective non-Abelian (Yang-Mills) gauge fields emerging in the vicinity of Fermi point, i.e. at low energy. For example, the Fermi point with $N=2$ may give rise to the $SU(2)$ gauge field $p_\beta^{(0)}=A_\beta^a\tau_a$, where 
$\tau_a$ are Pauli matrices corresponding to the emergent isotopic spin.

 \subsection{Emergent gravity} 

 The matrix field $e_i^k$ acts on the quasiparticles as  the field of vierbein, and thus describes the emergent dynamical gravity field. As a result, close to the Fermi point,  matter fields 
(all ingredients of Standard Model: chiral fermions and Abelian and non-Abelian gauge fields) 
emerge  together with geometry, relativistic spin, Dirac  matrices,  and physical laws:  Lorentz and gauge  invariance, equivalence principle, etc.
 In this scheme gravity emerges together with matter (Fig. \ref{EmergentPhysics}).   This means that the so-called  ``quantum gravity'' should be the unified theory of the underlying quantum vacuum, where the gravitational degrees of freedom cannot be separated from all other microscopic  degrees of freedom, which give rise to the matter fields (fermions and gauge fields).

Classical gravity would be a natural macroscopic phenomenon emerging in the low-energy corner of the microscopic quantum vacuum, i.e. it is a typical and actually inevitable consequence of the top (high energy) to bottom (low energy) coarse graining procedure. The inverse bottom to top procedure, i.e. from the classical to quantum gravity, is highly restricted.  The first steps in the quantization are certainly allowed: it is possible for example to quantize gravitational waves to obtain their quanta -- gravitons, since in the low energy corner the results of microscopic and effective theories coincide. It is also possible to obtain some (but not all) quantum corrections to Einstein equation; to extend classical gravity to the semiclassical and stochastic (Hu 2007) levels,  etc.  But one cannot  cannot  obtain ``quantum gravity'' by full quantization of Einstein equations, since all other degrees of freedom of quantum vacuum will be missed.

 \subsection{Dimension of space and flatness of Universe} 
 
  In the Fermi point scenario, space-time is naturally  4-dimensional.  This is the property of the Fermi-point topology, which as distinct from the string theory does not require the higher-dimensional space-times. 
 
 The Universe is naturally flat. In fundamental general relativity, the isotropic and homogeneous Universe means that the 3D space has a constant curvature. In emergent gravity with effective metric, the isotropic and homogeneous Universe corresponds to the flat 3D space. In general relativity the flatness of the  Universe requires either fine tuning or  inflationary scenario in which the curvature term is exponentially suppressed if the exponential inflation of the Universe irons out curved space to make it extraordinarily flat. The observed flatness of our Universe is in favor of emergent gravity. 

The effective gravity emerging at low energy may essentially differ from the fundamental
gravity even in principle. Since in the effective gravity the general covariance is lost at high energy, the
metrics which for the low-energy observers look as equivalent,
since they can be transformed to each other by coordinate
transformation, are not equivalent physically. As a result, some metrics, which are natural in general relativity, are simply forbidden in emergent gravity. 
For example, emergent gravity is not able to incorporate the geodesically-complete Einstein Universe with spatial section $S^3$  (Klinkhamer \& Volovik  2005$c$).  It, therefore, appears that the original static $S^3$ Einstein Universe  can exist only within the context of fundamental general relativity. 

In addition, some coordinate transformations in GR are not allowed in emergent gravity: these are either singular transformations of the original  coordinates, or the transformations which remove some parts of spacetime (or add the extra parts).
The non-equivalence of different metrics is especially important
in the presence of the event horizon. 
For example, in the emergent gravity the Painlev\'e-Gullstrand metric is more appropriate for the description of a black hole, than the Schwarzschild metric which is singular at the horizon.

 \section{Vacuum energy and  cosmological constant} 
\label{sec:CC}

There is a  huge contribution to the vacuum energy density of order
$E^4_{\rm P} \approx \big(10^{28}\,\text{eV}\big)^4$,
which comes from the ultraviolet  degrees of freedom,
whereas the observed
total energy density of approximately $\big(10^{-3}\,\text{eV}\big)^4$
is  smaller by many orders of magnitude. 
  In general relativity, the cosmological constant  is arbitrary constant, and thus its smallness requires fine-tuning. Thus observations are in favor of emergent gravity. 
If gravitation would be a truly fundamental interaction,
it would be hard to understand why the energies stored
in the quantum vacuum would not gravitate
at all (Nobbenhuis 2006).
If, however, gravitation would be only a low-energy effective interaction,
it could be that the corresponding gravitons as quasiparticles
do not feel \emph{all} microscopic degrees of freedom
(gravitons would be analogous to small-amplitude waves
at the surface of the ocean)
and that the gravitating effect of the vacuum energy density would be
effectively \emph{tuned away} and cosmological constant would be naturally small or zero  (Dreyer 2006). 

\subsection{Vacuum as self-sustained medium} 

A particular mechanism of nullification of the relevant vacuum energy works for such vacua which have the property of a \emph{self-sustained medium}.
A self-sustained vacuum is a medium with a definite macroscopic
volume even in the absence of an environment. An
example is a droplet of quantum liquid at zero temperature falling in empty space.
The observed near-zero value of the cosmological constant
compared to Planck-scale values 
suggests that the quantum vacuum of our universe
belongs to this class of systems.
As any other medium of this kind, the equilibrium vacuum
would be homogeneous and extensive. The homogeneity assumption is
indeed supported by the observed flatness and
smoothness of our universe (de Bernardis 2000, Hinshaw 2007, Riess 2007).
The implication is that the energy of the equilibrium quantum
vacuum would be proportional to the volume considered.

Usually, a self-sustained medium is characterized by an
\emph{extensive conserved quantity}
whose total value determines the actual volume of the
system (Landau \& Lifshitz 1980, Perrot 1998).
The Lorentz invariance of the vacuum imposes strong constraints on the possible form this variable can take.
One may choose the vacuum variable to be a
symmetric tensor $q^{\mu\nu}$ satisfying the following conservation law:
\begin{equation}
\nabla_\mu \,q^{\mu\nu} =0~,
\label{eq:conservation}
\end{equation}
In a homogeneous vacuum, one has $q^{\mu\nu}=q\,g^{\mu\nu}$ with $q$
constant in space and time.
The quantum vacuum can now be considered
as a reservoir of trans-Planckian energies stored in the $q$--field.

   \begin{figure}
\centerline{\includegraphics[width=0.8\linewidth]{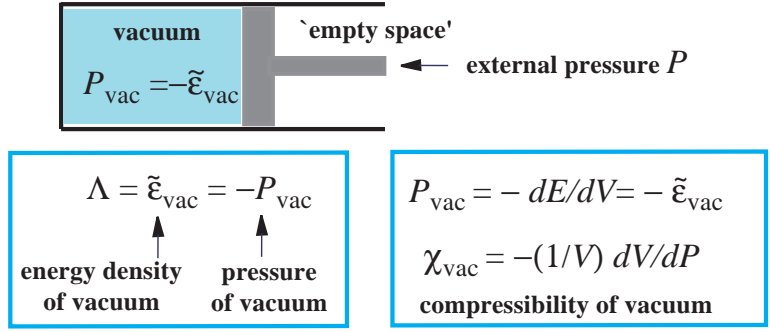}}
\caption{Vacuum as a medium obeying macroscopic thermodynamic laws. Relativistic vacuum possesses energy density, pressure and compressibility but has no momentum.  In equilibrium, the vacuum pressure $P_{\rm vac}$ equals the external pressure $P$ acting from the environment. The thermodynamic energy density of the vacuum $\tilde\epsilon_{\rm vac}$ which enters the vacuum equation of state $\tilde\epsilon_{\rm vac}= -P_{\rm vac}$ does not coincide with the microscopic vacuum energy $\epsilon$. While the natural value of  $\epsilon$ is determined by the Planck scale, $\epsilon\sim E_{\rm P}^4$,  the natural value of the macroscopic quantity $\tilde\epsilon$ is zero for the self-sustained vacuum which may exist in the absence of environment, i.e. at $P=0$. This explains why the present cosmological constant $\Lambda=\tilde\epsilon_{\rm vac}$ is small (Klinkhamer \& Volovik 2007)}  
\label{compressibility} 
\end{figure}

Let us consider  a large portion of quantum vacuum under external pressure $P$
(Fig. \ref{compressibility}). The volume $V$ of quantum vacuum is variable, but its total ``charge''
$Q(t)\equiv \int d^3r~q(\mathbf{r},t)$ must be conserved,
$\mathrm{d}Q/\mathrm{d}t=0$. 
The energy of this portion of quantum vacuum at fixed  total``charge'' $Q=q\, V$
is then given by the thermodynamic potential
\begin{equation}
W=E+P\,V=\int d^3r~\epsilon\left(Q/V\right) + P\,V~,
\label{eq:ThermodynamicPotential}
\end{equation}
where
$\epsilon\left(q\right)$ is the energy density in terms of $q$.
As the volume of the system is a free parameter,
the equilibrium state of the system is obtained by variation over the volume $V$:
\begin{equation}
\frac{d W}{dV}=0~,
\label{eq:Equilibrium}
\end{equation}
This gives an integrated form of the Gibbs--Duhem equation for the vacuum pressure:
\begin{equation}
P_{\rm vac}=-\epsilon(q) +q\,\frac{d\epsilon(q)}{dq}~,
\label{eq:Gibbs-Duhem}
\end{equation}
whose solution determines the equilibrium value $q=q_0(P)$
and the corresponding volume  $V=V_0(P,Q)=Q/q_0(P)$.

\subsection{Microscopic vs macroscopic vacuum energy} 

Equation (\ref{eq:Gibbs-Duhem}) suggests that the relevant thermodynamic potential of the vacuum energy, which is experienced by the low-energy degrees of freedom is:
\begin{equation}
\tilde\epsilon_{\rm vac}=\epsilon(q) -q\,\frac{d\epsilon(q)}{dq}~.
\label{eq:vev}
\end{equation}
This is confirmed by the example of the dynamic $q$ field, which demonstrates that the energy-momentum tensor of the vacuum is (Klinkhamer \&  Volovik 2007):
\begin{equation}
T_{\mu\nu}= g_{\mu\nu}\,\tilde\epsilon_{\rm vac}\,.
\label{eq:emSolution}
\end{equation}  
It is thus $\tilde\epsilon\left(q\right)$ rather than $\epsilon\left(q\right)$, which enters the equation of state for the vacuum and thus corresponds to the cosmological constant:
\begin{equation}
\Lambda=\tilde\epsilon_{\rm vac}=-P_{\rm vac}~.
\label{eq:EoS}
\end{equation}

\subsection{Natural value of cosmological constant} 

 \begin{figure}
\centerline{\includegraphics[width=0.9\linewidth]{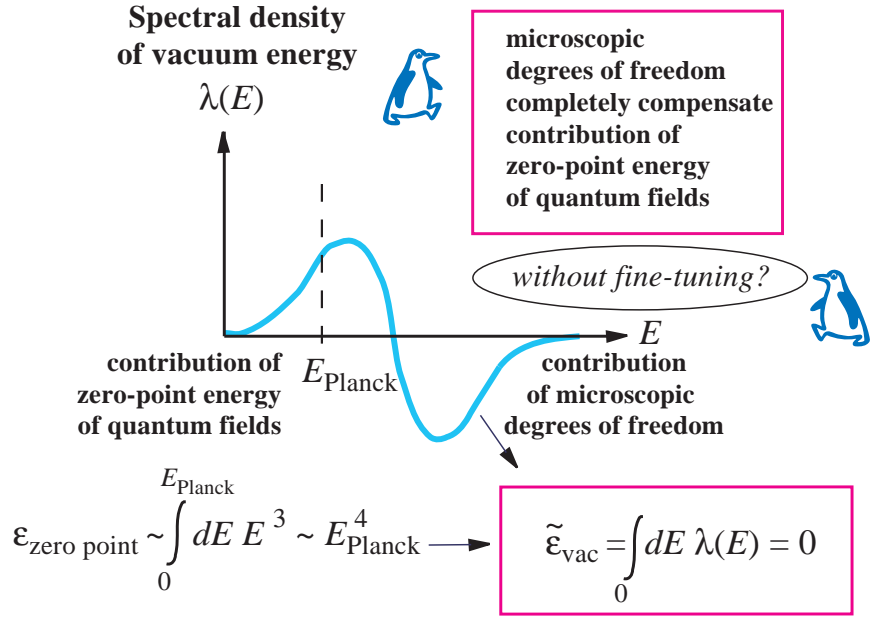}}
\caption{Contribution of different energy scales into the energy of the self-sustained vacuum. Zero point energy of bosonic fields gives rise to the diverging contribution to the energy  of quantum vacuum, which is compensated by microscopic (trans-Planckian) degrees of freedom. }  
\label{spectrum} 
\end{figure}

While the energy of microscopic quantity $q$ is determined by the Planck scale,
$\epsilon(q_0) \sim E_{\rm P}^4$, the real vacuum energy which sources the effective gravity is determined by a macroscopic quantity -- the external pressure.
In the absence of an environment, i.e. at zero external pressure, one obtains that the pressure of pure and equilibrium vacuum is exactly zero:
\begin{equation}
\Lambda=-P_{\rm vac}=-P=0~.
\label{eq:Null}
\end{equation}
Thus from the thermodynamic arguments it follows that for any effective theory of gravity the natural value of $\Lambda$ is zero. This result does not depend on the microscopic structure of the vacuum  from which gravity emerges, and is actually the final result of the renormalization dictated by macroscopic physics. In the self-sustained vacuum, the huge contribution of zero-point energy of macroscopic fields to the vacuum energy $\tilde\epsilon_{\rm vac}$ is compensated by microscopic degrees of the vacuum
(Volovik 2008, see Fig. \ref{spectrum}). If the cosmological phase transition takes place, the vacuum is readjusted to a new equilibrium and $\Lambda$ approaches zero again  (Klinkhamer \& Volovik 2007).

 \subsection{Compressibility of the vacuum}

Using the standard definition of the inverse of
the isothermal compressibility, $\chi^{-1} \equiv -V\,dP/dV$ (Fig. \ref{compressibility}),  one obtains the compressibility of the vacuum by varying Eq.(\ref{eq:Gibbs-Duhem}) at fixed $Q=qV$:
\begin{equation}
\chi_\text{vac}^{-1} \equiv -V\frac{dP_{\rm vac}}{dV}=\left[q^2\;\frac{d^2\epsilon(q)}{dq^2}\,\right]_{q=q_0}
> 0~.
\label{eq:Stability}
\end{equation}
A positive value of the vacuum compressibility
is a necessary condition for the stability of the vacuum. It is, in fact, the stability of
the vacuum, which is at the origin of the nullification of the cosmological
constant 
in the absence of an external environment.

From the low-energy point of view, the compressibility of the vacuum $\chi_\text{vac}$ is as fundamental physical constant as the Newton constant $G$, although $\chi_\text{vac}$ is not yet observed.  While the natural value of the macroscopic quantity $P_{\rm vac}$ (and $\Lambda$) is zero, the natural values of  the parameters $G$ and $\chi_\text{vac}$ are determined by the Planck physics and are expected to be of order $1/E^{2}_{\rm P}$ and $1/E^{4}_{\rm P}$ correspondingly (Table 1).

 \subsection{Thermal fluctuations of $\Lambda$ and the volume of Universe}

The compressibility of the vacuum $\chi_\text{vac}$, though not measurable at the moment, can be used for estimation of  the lower limit for the volume $V$ of the Universe. This estimation follows from the upper limit for thermal fluctuations of cosmological constant (Volovik 2004). The mean square of thermal  fluctuations of $\Lambda$ equals the mean square of thermal  fluctuations of the vacuum pressure, which in turn is determined by thermodynamic equation (Landau \& Lifshitz 1980):
\begin{equation}
\left <\left(\Delta\Lambda\right)^2\right>=\left <\left(\Delta P\right)^2\right>=\frac{ T}{V\chi_{\rm vac}}~.
\label{Fluctuations}
\end{equation}
 Typical fluctuations of the cosmological constant $\Lambda$ should not exceed the observed value:  $\left <\left(\Delta\Lambda\right)^2\right>< \Lambda_{\rm obs}^2$. Let us assume, for example, that the temperature of the Universe is determined by the temperature $T_{\rm CMB}$ of the cosmic microwave background radiation. Then, using our estimate for vacuum compressibility $\chi_{\rm vac}^{-1}\sim E^4_{\rm P}$, one obtains that the volume $V$ of our Universe highly exceeds the Hubble volume $V_H=R_H^3$ --  the volume of visible Universe inside the present cosmological horizon:
\begin{equation}
V> \frac{T_{\rm CMB}} { \chi_{\rm vac} \Lambda_{\rm obs}^2}\sim 10^{28}V_H~.
\label{Volume}
\end{equation}
This demonstrates that the real volume of the Universe is certainly not limited by the present cosmological horizon.

\section{Energy scales and physical laws} 

\subsection{Hierarchy of energy scales} 

 \begin{figure}
\centerline{\includegraphics[width=1.0\linewidth]{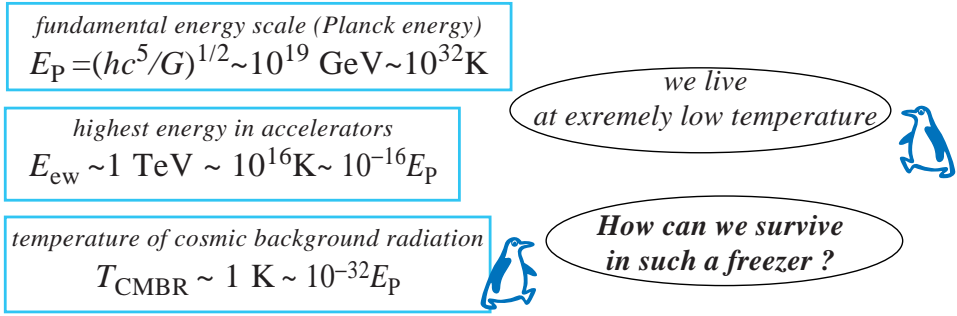}}
\caption{Characteristic high-energy scale in the vacuum of the ``natural Universe''
 is the Planck energy $E_P$. Compared to that energy, the high-energy physics and cosmology operate at extremely ultra-low temperatures.}  
\label{MainProblem} 
\end{figure} 

\begin{figure}
\centerline{\includegraphics[width=1.0\linewidth]{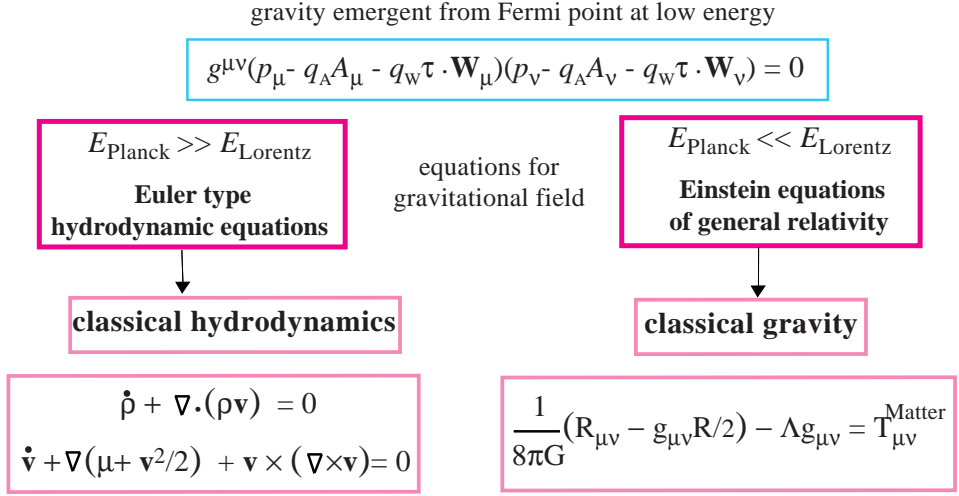}}
\caption{Equations for the metric field  $g_{\mu\nu}$ emerging near the Fermi point depend on hierarchy of ultraviolet cut-off's:
Planck energy scale $E_{\rm P}$ vs Lorentz violating scale $E_{\rm Lorentz }$.  Classical Einstein equations for $g_{\mu\nu}$ emerge only in the limit when the Lorentz invariance is  fundamental at the Planck scale, i.e. when  $E_{\rm Lorentz} \gg E_{\rm P}$}
\label{GravityAndHierarchy} 
\end{figure}

In the emergent physics, the energy scale in our Universe is not limited by the characteristic Planck scale.  To obtain the observed high precision of physical laws, the Lorentz symmetry must persist well above the Planck energy. This requirement  represents the most crucial test for the emergent scenario.  In the case when the Lorentz violating scale $E_{\rm Lorentz }< E_{\rm P}$ the metric field does not obey Einstein equations; instead it is governed by the hydrodynamic type equations (see Fig. \ref{GravityAndHierarchy}). The Einstein equations emerge in the limit $E_{\rm Lorentz }\gg E_{\rm P}$. This can be seen on the example of the Frolov-Fursaev version of Sakharov induced gravity (Frolov and Fursaev 1998), where the ultraviolet cut-off is much larger than the Planck energy, and Einstein equations are reproduced. Accuracy of Einstein equations is determined by the small parameter $E_{\rm P}^2/E_{\rm Lorentz }^2\ll 1$. The same parameter would enter the mass of the emergent gauge bosons, $M\sim E_{\rm P}^2/E_{\rm Lorentz }$  (Klinkhamer \& Volovik 2005$a$, see Table 1). Experimental bounds on the violation of Lorentz symmetry can be  obtained from ultra-high-energy cosmic rays.   For example,  according to conservative estimates, the relative value of  the Lorentz violating terms in Maxwell equation is smaller than $10^{-18}$ (Klinkhamer \& Risse 2008). This suggests that  $E_{\rm Lorentz } > 10^9E_{\rm P}$.

 All this implies that physics continues far beyond the Planck scale, and this opens new  possibilities for construction of microscopic theories.  Since in the Fermi point scenario bosons are the composite objects, the ultraviolet  cut-off may be different for fermions and bosons (Klinkhamer \& Volovik 2005$a$). The smaller (composite) scale can be associated with $E_{\rm P}$, while
the "atomic" structure of the quantum vacuum will be  revealed only at the
much higher Lorentz-violating scale $E_{\rm Lorentz}$. 
 
 A first step towards the elusive theory of ``quantum gravity'' would be to identify the microscopic constituents (`atoms') of space.
At the moment we are not able to do this, but we can estimate the number of the underlying  Òatoms of the
etherÓ, whatever they are. This is the volume of our Universe
within the cosmological horizon divided by the elementary Planck volume:
$N\sim R_H^3/l_{\rm Planck}^3\sim 10^{180}$. At least an extra 30 orders of magnitude must be added if the real volume  of the Universe in Eq.(\ref{Volume}) is considered: $N\sim V/l_{\rm Planck}^3\sim 10^{210}$. 
Finally if  we relate the minimum  length to  the   Lorentz-violating scale $l_{\rm Lorentz}$ and take 
 $E_{\rm Lorentz}\sim 10^{20} E_{\rm Planck}$, we obtain $N\sim V/l_{\rm Lorentz}^3\sim 10^{270}$
constituents.
 
The large number $N$ of the constituents of a system means that the system
is macroscopic and must obey the laws of  emergent macroscopic physics. The most general physical laws which do not depend on the details of
the underlying microscopic system are the laws of  thermodynamics. 
 The huge number $N$ of the constituents could lead to the hierarchical structure of physics,
 with different physical laws  emerging at different levels of the hierarchical structure
 with highly separated length scales: 
 \begin{equation}
l_{\rm Lorentz}\ll l_{\rm P} \ll l_{\rm ew} \ll l_{\rm QCD} \ll \ldots \ll R_H \ll V^{1/3}~.
\label{hierarchy}
\end{equation}
 The accuracy of the physical laws at a given scale $l_n$ is determined by the parameter $l_{n-1}/l_n \ll 1$. If this parameter is not very small,  the physical laws at scale $l_n$ are rather crude and contain a lot of phenomenological parameters coming from the smaller length scales.
 
Simple and accurate physical laws may emerge only for a big number of constituents and/or for the large ratio between the adjacent scales (Bjorken 2001).

 \subsection{Natural values of physical quantities}

Both in particle physics and condensed matter the natural value of a quantity depends on whether this quantity is determined by macroscopic or  microscopic physics, see Table 1.

    \begin{table}
    \label{Table1}
   \caption{Natural values of physical quantities}
   \longcaption{Natural values of physical quantities following from the  Fermi point (FP) scenario of emergent physics vs naive estimates and observation. Here $E_{\rm UV}\gg E_{\rm P} \gg E_{\rm IR} \gg R_H^{-1}$ 
   are, respectively, the ultra-violet (Lorentz violating?) energy scale, the Planck energy scale, the infra-red cut-off and the inverse Hubble radius}
   \begin{tabular}{lccc}
 \hline
 \hline
    \cr
  {\rm  physical~quantity} &~~{\rm naive~estimate}  ~~&~~{\rm natural~FP~value} ~~ &~~{\rm observed~value}   
   \cr
   \cr
\hline
\hline
   \cr
     {\rm mass~of~elementary~particle}     & $E_{\rm P} $     &0            &$\lll E_{\rm P} $     
    \cr
     {\rm mass~of~gauge~boson}         & $E_{\rm P} $   &$E_{\rm P}^2/E_{\rm UV} $           &$\lll E_{\rm P} $     
    \cr
  {\rm running ~coupling ~constants}  & $ \ln^{-1} (E_{\rm P}/E_{\rm IR})$ & $ \ln^{-1} (E_{\rm P}/E_{\rm IR})$     &$ \sim$ 1 
   \cr
  {\rm Newton ~constant}  &$E_{\rm P}^{-2} $ &$E_{\rm P}^{-2} $   &$E_{\rm P}^{-2} $  
    \cr
    \cr
\hline
\hline
    \cr
    {\rm temperature~of~Universe}  &$E_{\rm P} $  &0      &$\lll E_{\rm P} $  
    \cr
 {\rm cosmological ~constant}  &$E_{\rm P}^4$ &0 ~or~$\sim E_{\rm P}^2/R_H^2$  &$\sim E_{\rm P}^2/R_H^2$ 
  \cr
 {\rm volume~of~Universe}          &$E_{\rm P}^{-3}$~ or~$R_H^{3}$&$\gg R_H^{3}$            &$\gg R_H^{3}$     
 \cr
  {\rm curvature~of~Universe}       &$E_{\rm P}^{2}$~  or ~$R_H^{-2}$     &0        &$\ll R_H^{-2}$    
 \cr
  {\rm vacuum~compressibility}       &$E_{\rm P}^{-4}$   &$E_{\rm P}^{-4} $            &$-$     
  \cr
  \cr
\hline
\hline
   \end{tabular}
 \end{table}

   The first column in the Table 1 contains naive estimates of physical quantities. They follow from the dimensional analysis assuming that the role of the fundamental scale is played either by Planck energy  $E_{\rm P} $ or by the size of Universe $R_H$.  The second column shows the  natural values of these quantities which follow from the Fermi point scenario.  In most cases the naive estimate contradicts  both to the values dictated by the Fermi point scenario and to observations shown in the third column.
 
The naive estimates are consistent with  natural values  for those quantities which are determined by microscopic physics and are expressed in terms of the corresponding microscopic scale, which is the Planck scale $E_{\rm P}$ in our Universe or atomic scale in condensed matter systems.  Example  is the Newton constant $G=a_G E_{\rm P}^{-2}$. For emergent gravity, the dimensionless prefactor $a_G$  depends on the vacuum content and is of order unity in units $\hbar=c=1$. In principle, the parameter  $a_G$ can be zero, but this requires the fine-tuning between different scalar, vector and spinor fields in the vacuum. That is why the natural value of $G$ is $E_{\rm P}^{-2}$. The compressibility of the vacuum is also determined by microphysics.
The running coupling constants $\alpha_n$ also fall into this category, since they depend on the ultraviolet cut-off together with the infra-red cut-off  $E_{\rm IR}$: $\alpha_n^{-1}\sim \ln (E_{\rm P}/E_{\rm IR})$.

Temperature,  pressure  and the volume of Universe belong to the category determined by macroscopic physics -- thermodynamics. These thermodynamic quantities do not depend on the micro-physics or on momentum-space  topology; they only depend on the environment. In the absence of the forces from the environment, the pressure and temperature of any system relax to zero. The same should hold for the temperature of the Universe and for the vacuum pressure. The vacuum pressure is with the minus sign the cosmological constant, $\Lambda=\epsilon_{\rm vac}=-p_{\rm vac}$. Whatever is the vacuum content, and independently of the history of the phase transitions in the quantum vacuum, the cosmological constant must relax to zero or to the small value which compensates the other partial contributions to the total pressure of the system, since it is the total pressure of the system that must be zero in equilibrium in the absence of the environment.

 \subsection{Mass is naturally zero}

Masses of elementary fermions are quantities which most crucially depend on the momentum space topology. The naive estimation tells us that these masses should be on the order of Planck energy scale:  $M_{\rm theor} \sim E_{\rm P}\sim 10^{19}$ GeV. This highly contradicts to observations: the observed masses of known particles are many orders of magnitude smaller being below the electroweak energy scale $M_{\rm obs}<E_{\rm ew}\sim 1$  TeV.  This  represents the main hierarchy problem. In the ``natural''   Universe, where all masses are of order $E_{\rm P}$, all fermionic degrees of freedom are completely frozen out because of the  Bolzmann factor $e^{-M/T}$, which is about $e^{-E_{\rm P} /E_{\rm ew}} \sim e^{-10^{16}}$  already at the temperature corresponding to the  highest energy reached in accelerators. There is no  fermionic matter in such a Universe.  

That we survive in our Universe is not the result of the anthropic principle (the latter chooses the Universes which are fine-tuned for life but have an extremely low probability). On the contrary, this simply indicates that our Universe is also natural, and its vacuum is generic though belongs to a different universality class of vacua --  the vacua with Fermi points. In such vacua the masslessness of fermions is protected by topology (combined with symmetry, see Sec. \ref{SymmetryTopology}).

As for masses of gauge bosons, they may appear either due to symmetry breaking occurring at low energy, or due to the higher order  corrections to the effective action emergent close to the Fermi point. In the latter case, the mass is determined by the hierarchy of scales $E_{\rm P}$ and $E_{\rm UV}$, say by $E_{\rm P}^2/E_{\rm UV}$, where the higher energy scale may correspond to the  Lorentz violating scale $E_{\rm Lorentz }$ (Klinkhamer \& Volovik 2005$a$).

 \section{Symmetry vs topology} 
 \label{SymmetryTopology}

 \subsection{Discrete symmetries in the underlying physics}

\begin{figure}
%\centerline{\includegraphics[width=1.0\linewidth]{SymVsTopology.eps}}
\centerline{\includegraphics[width=1.0\linewidth]{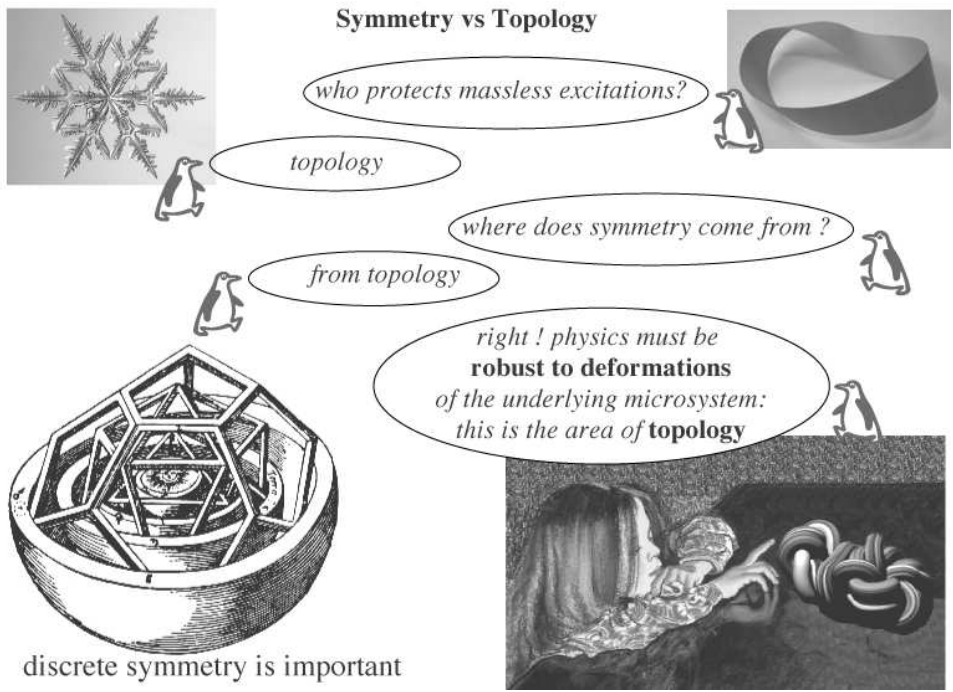}}
\caption{Momentum-space topology is the main source of massless elementary paprticles. But it must be accompanied by discrete symmetries between Fermi points (see Fig. \protect\ref{TwoScenarios}). {\it bottom right}: from ``Knots in art''  by  Piotr Pieranski. }  
\label{SymVsTopology.eps}
\end{figure}

Standard Model above the electroweak transition contains 16 chiral fermions in each
generation: 8 right-handed fermions with topological charge $N=+1$ each and 8 left-handed
fermions with $N=-1$ each. The vacuum of the Standard
Model
above the electroweak transition
is marginal: there is a multiply
degenerate Fermi point at ${\bf p}=0$ with the total topological charge
$N=+8-8=0$.

The absence of the topological
stability means that even  a small mixing between
the fermions may lead to annihilation of the marginal Fermi point.
 In the Standard Model, the proper mixing which leads to the fully gapped
vacuum  is prohibited by  two discrete symmetries: electroweak $Z_2$ symmetry  (Volovik 2003) and CPT.

\begin{figure}
\centerline{\includegraphics[width=0.7\linewidth]{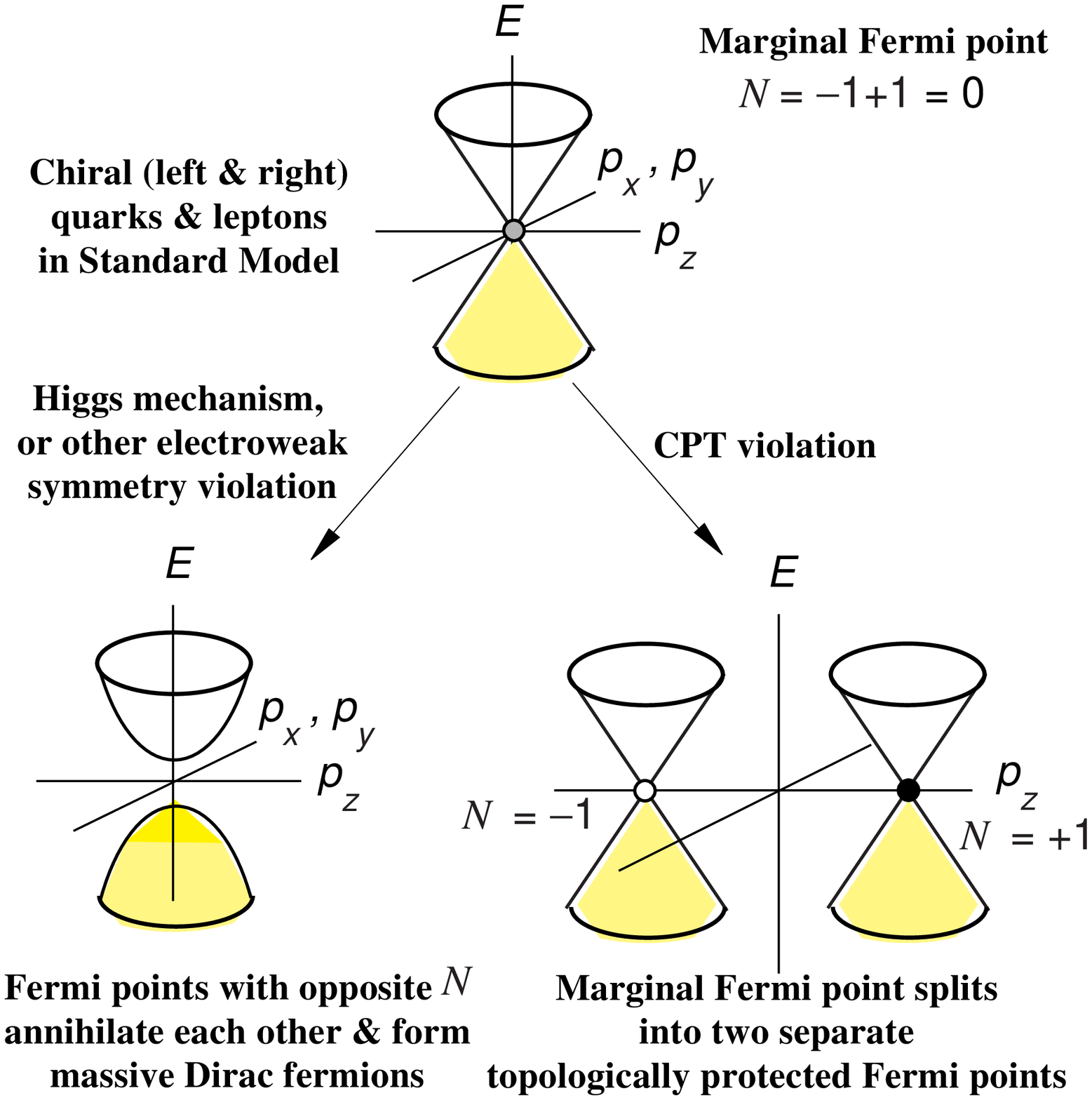}}
\caption{({\it top}) In Standard Model the Fermi points with positive $N=+1$ and negative $N=-1$ topological charges are at the same point ${\bf p}=0$. It is the discrete symmetry between the Fermi points which prevents their mutual annihilation. When this symmetry is violated or spontaneously broken, there are two topologically different scenarios: ({\it bottom left}) either Fermi point annihilate each other and  Dirac mass is formed;  ({\it bottom right})  or Fermi points split
(Klinkhamer \& Volovik  2005$b$). It is possible that actually the splitting exists at the microscopic level,  but in our low energy corner we cannot observe it because of the emergent gauge symmetry: in some cases  splitting can be removed by  gauge transformation.}  
\label{TwoScenarios} 
\end{figure}

This means that underlying physics must contain discrete symmetries (Fig. \ref{SymVsTopology.eps}). Their role is extremely important. The main role is to prohibit the cancellation of the Fermi points with opposite topological charges (see Fig. \ref{TwoScenarios}). As a side effect, in the low-energy corner discrete symmetries are transformed into gauge symmetries and give rise to effective non-Abelian gauge field. In particular, the $Z_2$ symmetry produces the $SU(2)$ gauge field (Volovik 2003). Discrete symmetries also reduce the number of the massless gauge bosons and the number of metric fields. To justify the Fermi point scenario, one should find such discrete symmetry which leads in the low energy corner to one of the GUT or Pati-Salam models.

 \subsection{Discrete symmetries and splitting of Fermi points}

Explicit violation or spontaneous breaking of one of the two discrete symmetries 
transforms the marginal vacuum of the Standard Model
into one of the two topologically-stable vacua.  If, 
for example, the electroweak $Z_2$ symmetry is broken, the marginal Fermi point
disappears and the fermions become massive  
[Fig.~\ref{TwoScenarios})]. This is assumed to happen below the
symmetry breaking electroweak transition caused by Higgs mechanism
where quarks and charged leptons acquire the Dirac masses.  
If, on the other hand, the CPT symmetry is violated, the
marginal Fermi point splits into topologically-stable Fermi
points which protect massless chiral fermions.  One can speculate that in the  Standard Model the latter happens
with the electrically neutral leptons, the neutrinos
(Klinkhamer \& Volovik 2005$b$). Most interestingly, Fermi-point splitting of neutrinos
may provide a new source of T and CP violation
in the leptonic sector, which may be relevant
for the creation of the observed cosmic
matter-antimatter asymmetry (Klinkhamer 2006).
Examples of discrete symmetry and splitting of Fermi and Majorana points in condensed matter are discussed in the review paper by Volovik (2007).

 \begin{figure}
%\centerline{\includegraphics[width=1.0\linewidth]{Emergent-Breaking_paper.eps}}
\centerline{\includegraphics[width=1.0\linewidth]{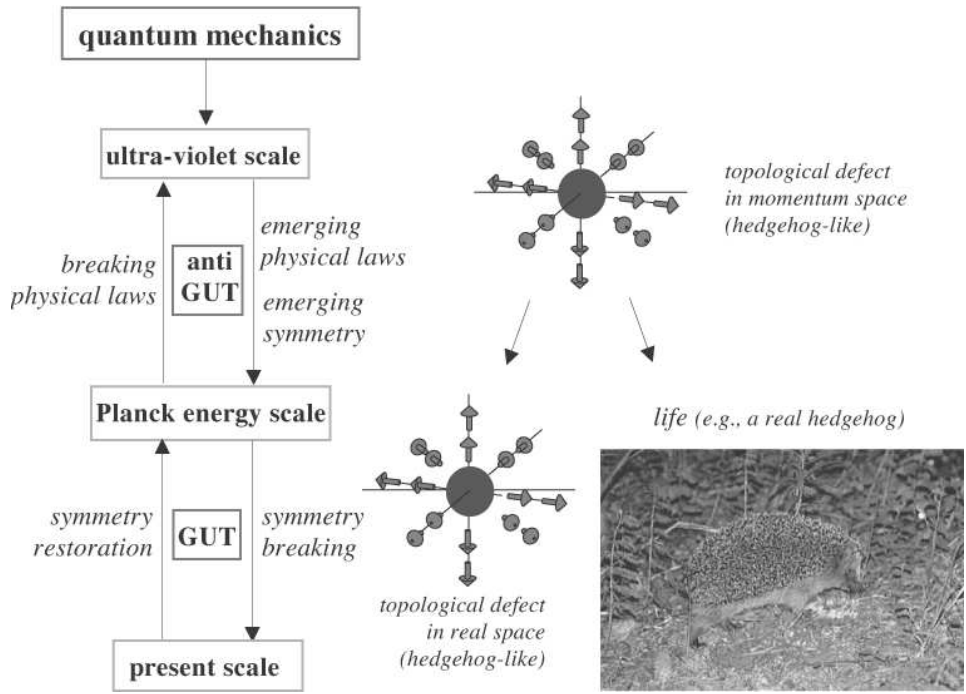}}
\caption{Three elements of modern physics: (i) quantum mechanics (or quantum field theory); (ii) Grand Unification based on the phenomenon of broken symmetry at low energy (GUT symmetry is restored when the Planck energy scale is approached from below); and (iii) anti-GUT based on the opposite phenomenon of emergent symmetry -- the GUT symmetry gradually emerges when the Planck energy scale is approached from above.  A hedgehog-like  topological defect  in momentum space -- the Fermi point -- gives rise to symmetry emergent  at Planck-GUT scales. At lower energy  the GUT symmetry is broken giving rise to topological defects in real space (e.g., a hedgehog-like object)  and life (e.g., a real hedgehog).}  
\label{TriKita} 
\end{figure}

 \section{Conclusion} 
 
 There are two complementary schemes for the classification of quantum vacua,  both are based on quantum mechanics which  is assumed to be a fundamental theory (Fig. \ref{TriKita}). The traditional classification -- the Grand Unification  (GUT)  scheme --  assumes that fermionic and bosonic  fields and gravity are also the fundamental  phenomena. They  obey the fundamental symmetry which becomes spontaneously broken at low energy, and is restored when the Planck energy scale is approached from below.  
 
 The Fermi point scenario provides a complementary anti-GUT scheme in which the `fundamental' symmetry and `fundamental'  fields of GUT  gradually emerge together with `fundamental' physical laws when the Planck energy scale is approached from above. The emergence of the `fundamental'  laws of physics is provided by the general property of topology -- robustness to details of the microscopic trans-Planckian physics. In these scheme, fermions are primary objects. Approaching the Planck energy scale from above, they are transformed to the Standard Model chiral fermions  and  give rise to the secondary objects: gauge fields and gravity. Below the Planck scale,  the GUT scenario  intervenes giving rise to symmetry breaking at low energy. This is accompanied by formation of  composite objects, Higgs bosons,
 and tiny Dirac masses of quark and leptons.
 
In the GUT scheme, general relativity is assumed to be as fundamental as quantum mechanics, while in the second scheme  general relativity  is a secondary phenomenon. In the anti-GUT scheme, general relativity is the effective theory describing the dynamics of the effective metric experienced by the effective low-energy fields. It is a side product of quantum field theory or of the quantum mechanics in the vacuum with Fermi point. As a result, in the Fermi-point scenario there is no principle contradictions  between quantum mechanics and gravity. That is why, the emergent gravity cannot be responsible for the issues related to foundations of quantum mechanics, and in particular for the collapse of the wave function. Also, the hierarchy of scales implies that if  quantum mechanics is not fundamental, the scale at which it emerges should be far above the Planck scale.

    \begin{acknowledgements}

I thank Frans Klinkhamer for fruitful discussions. This work has been supported in part  by the Russian
Foundation for Fundamental Research  (grant 06--02--16002--a). 
 \end{acknowledgements}

\end{document}